\documentclass[manuscript,screen,review，nonacm]{acmart}
\AtBeginDocument{%
  \providecommand\BibTeX{{%
    \normalfont B\kern-0.5em{\scshape i\kern-0.25em b}\kern-0.8em\TeX}}}

\setcopyright{acmcopyright}
\copyrightyear{2023}
\acmYear{2023}
\acmDOI{XXXXXXX.XXXXXXX}

\acmConference[xxxxxx]{Make sure to enter the correct
  conference title from your rights confirmation emai}{xxxxxxxx}{xxxxxxxx}
\acmBooktitle{xxxxxxxxxxxxxxxxxxxxxxxxxx} 
\acmPrice{15.00}
\acmISBN{978-1-4503-XXXX-X/18/06}

\usepackage{algorithmic}
\usepackage{algorithm}
\usepackage{subfigure}
\usepackage{booktabs}

\begin{document}

\title{GNN4FR: A Lossless GNN-based Federated Recommendation Framework}


\author{Guowei Wu}
\affiliation{%
	\institution{College of Computer Science and Software Engineering, Shenzhen University}
	\city{Shenzhen}
	\country{China}
}
\email{2019111045@email.szu.edu.cn}

\author{Weike Pan}
\affiliation{%
	\institution{College of Computer Science and Software Engineering, Shenzhen University}
	\city{Shenzhen}
	\country{China}
}
\email{panweike@szu.edu.cn}
\authornote{co-corresponding authors}

\author{Zhong Ming}
\affiliation{%
	\institution{College of Computer Science and Software Engineering, Shenzhen University}
	\city{Shenzhen}
	\country{China}
}
\email{mingz@szu.edu.cn}
% \author{ANONYMOUS AUTHOR(S)}


\begin{abstract}
Graph neural networks (GNNs) have gained wide popularity in recommender systems due to their capability to capture higher-order structure information among the nodes of users and items. However, these methods need to collect personal interaction data between a user and the corresponding items and then model them in a central server, which would break the privacy laws such as GDPR. So far, no existing work can construct a global graph without leaking each user's private interaction data (i.e., his or her subgraph). In this paper, we are the first to design a novel lossless federated recommendation framework based on GNN, which achieves full-graph training with complete high-order structure information, enabling the training process to be equivalent to the corresponding un-federated counterpart. In addition, we use LightGCN to instantiate an example of our framework and show its equivalence.

\end{abstract}






\maketitle

\section{Introduction}
Recommender systems have played an important role in our lives, which are used to help users filter out the information they are not interested in. GNNs are widely used in personalized recommendation methods as they are able to capture high-order interactions between users and items in a user-item graph, enhancing user and item representations~\cite{berg2017graph,wang2019neural,wang2019kgat,fan2019graph,zhang2019star,ying2018graph}. However, these methods face challenges in terms of privacy laws, such as GDPR~\cite{voigt2017eu} as they require the collection and modeling of personal data in a central server.\par 
Constructing the global graph using all users’ subgraphs is often not allowed. Therefore, existing works~\cite{wu2022federated,qu2023semi} just expand a user's local graph to exploit high-order information. \par 
In this paper, we propose the first lossless federated framework named GNN4FR, which can accommodate almost all existing graph neural networks (GNNs) based recommenders. The contributions of this paper are summarized as follows:
% (i)We propose a novel lossless federated framework for GNN-based methods, which enables the training process to be equivalent to the corresponding un-federated counterpart;
% (ii)We propose an “expanding local subgraph + synchronizing user embedding” mechanism to achieve full-graph training;
% (iii)We choose LightGCN~\cite{he2020lightgcn} as an instantiation of our framework to demonstrate its equivalence.

\begin{itemize}
    \item We propose a novel lossless federated framework for GNN-based methods, which enables the training process to be equivalent to the corresponding un-federated counterpart.
    \item We propose an “expanding local subgraph + synchronizing user embedding” mechanism to achieve full-graph training.
    \item We choose LightGCN~\cite{he2020lightgcn} as an instantiation of our framework to demonstrate its equivalence.
\end{itemize}

\section{Related Work}

% \subsection{Federated Recommendation}
{\noindent \bf Federated Recommendation} Recommendation systems have seen significant growth in today's society. However, the training and inference processes of these models heavily rely on users' personal data, which raises concerns about privacy. With the introduction of GDPR~\cite{voigt2017eu}, the need for privacy and security in the recommendation domain led to the emergence of federated learning~\cite{yang2019federated}. This approach aims to address privacy issues through decentralized model training.
In the field of recommendation, several frameworks have been developed to enable federated learning~\cite{ammad2019federated,chai2020secure,flanagan2021federated,hegedHus2019decentralized,qi2020privacy,wu2022federated,qu2023semi,luo2022personalized,li2022fedgrec,liu2022federated}. For instance, FCF~\cite{ammad2019federated} and FedMF~\cite{chai2020secure} are specifically designed for factorization-based recommendation models. When it comes to GNN-based recommendation models, there are some existing frameworks such as FedGNN~\cite{wu2022federated} and SemiDFEGL~\cite{qu2023semi}. However, these frameworks are not able to construct a global graph without resorting to other entities or information, leading to some loss of the high-order structure information.

% \subsection{GNN-based Recommendation}
{\noindent \bf GNN-based Recommendation} Graph neural networks (GNNs) have gained wide popularity in recommender systems due to their capability to capture higher-order information. Notable examples include NGCF~\cite{wang2019neural} and LightGCN~\cite{he2020lightgcn}. NGCF employs a 3-hop graph neural network to learn user and item embeddings. Subsequently, LightGCN improves upon NGCF by eliminating redundant components and achieving superior results.
However, these methods are centralized and rely on collecting user information through the server to construct a global graph. Yet, privacy concerns make it challenging for the server to collect user data for graph construction. To address this limitation, we propose a novel framework that enables the server to construct the global graph using distributed user data, while ensuring user privacy protection. This approach achieves equivalent performance to centralized graph model training, making it the first lossless GNN-based federated recommendation framework to date.

\section{GNN4FR}

\begin{algorithm}[ht]
	\caption{GNN4FR}
        \label{GNN4FR}
	\begin{algorithmic}[1]
		\STATE Initialize(), i.e., Algorithm~\ref{Initialize}
		\STATE ExpandLocalGraph(), i.e., Algorithm~\ref{Expand the local subgraph}
		\FOR{ $ t=1, 2, \ldots, T$ }
		\STATE ForwardPropagation(), i.e., Algorithm~\ref{Train. forward propagation}
		\FOR{each client $u$ in parallel}
		\STATE constructs the local loss function
		\STATE computes the gradient of the local loss w.r.t. the nodes of the final layer	
		\ENDFOR	
		\STATE BackwardPropagation(), i.e., Algorithm~\ref{Train. backward propagation}
		\STATE AggregateGradients(), i.e., Algorithm~\ref{Train. aggregate the gradients}
		\STATE $\theta =\theta -\gamma \nabla_\theta$
		\ENDFOR
	\end{algorithmic}
\end{algorithm}
This section describes our proposed GNN4FR in detail, i.e., Algorithm~\ref{GNN4FR}. The training process contains five major parts. Firstly, do pre-training preparations, including initializing item embeddings, etc. Secondly, expand the subgraphs of all clients. Thirdly, use an “expanding local subgraph + synchronizing user embedding” mechanism for forward propagation. Fourthly, pass back neighboring users' embedding gradients. Fifthly, use secret sharing to aggregate gradients. In order to make it more comprehensive, we illustrate this process using a specific example to enhance clarity.\par 
In the example, there are three clients and four items, and their interactions are shown in Figure~\ref{fig_expand_subgraph}. Besides, the number of GNN convolution layers is three. 
\begin{algorithm}[ht]
	\caption{Initialize}
        \label{Initialize}
	\begin{algorithmic}[1]
		\FOR{each client $u$ in parallel}
		\STATE constructs the local subgraph $\mathcal{G}_u$
		\STATE initializes parameters (i.e., user embedding and item embeddings) using the same seed
		\STATE generates key pairs including the public key $PB_u$ and the private key $PI_u$
		\STATE sends $PB_u$ to the server
		\ENDFOR
		\STATE the server collects $PB_u$ from each client $u$, and sends to the client $u_c$ (i.e., the user randomly chosen by the server)
		\STATE the client $u_c$ receives the public keys from the server, encrypts the shared key $S$ with each $PB_u$, and sends them to the server
		\STATE the server receives the ciphertext and forwards them to the corresponding clients
		\FOR{each client $u$ in parallel}
		\STATE receives the ciphertext of $S$
		\STATE decrypts it with $PI_u$ and obtains the shared key $S$
		\ENDFOR		
	\end{algorithmic}
\end{algorithm}
For doing pre-training preparations, we show the process in Algorithm~\ref{Initialize}. Firstly, each client constructs his or her local subgraph and initializes the user embedding and item embeddings using the same seed, which means that the embedding of item $i_0$ in client $u_0$ is equal to the embedding of item $i_0$ in client $u_1$. Secondly, each client generates a key pair (i.e., a public key and a private key) and sends the public key to the server. Thirdly, the server collects the public keys of all clients, then randomly chooses one client $u_c$ from all clients (here we suppose that $u_c$ is $u_1$), and sends all the public keys to it (i.e., $u_1$). Fourthly, the client $u_1$ generates a shared key $S$, encrypts it with all the public keys received from the server (i.e., public keys of $u_2$ and $u_3$), and sends all the ciphertext to the server. Fifthly, the server receives all the ciphertext and forwards them to the corresponding clients. Finally, each client receives the ciphertext of the shared key $S$ and decrypts it with his or her own private key. This means that the clients $u_1$, $u_2$, and $u_3$ would hold the common shared keys $S$, but the server did not know it.\par

\setcounter{figure}{-1} % 设置图的计数器为-1

\begin{figure}
\centering
\subfigure{
\begin{minipage}[t]{0.49\linewidth}
\centering
\includegraphics[width=1.0\linewidth]{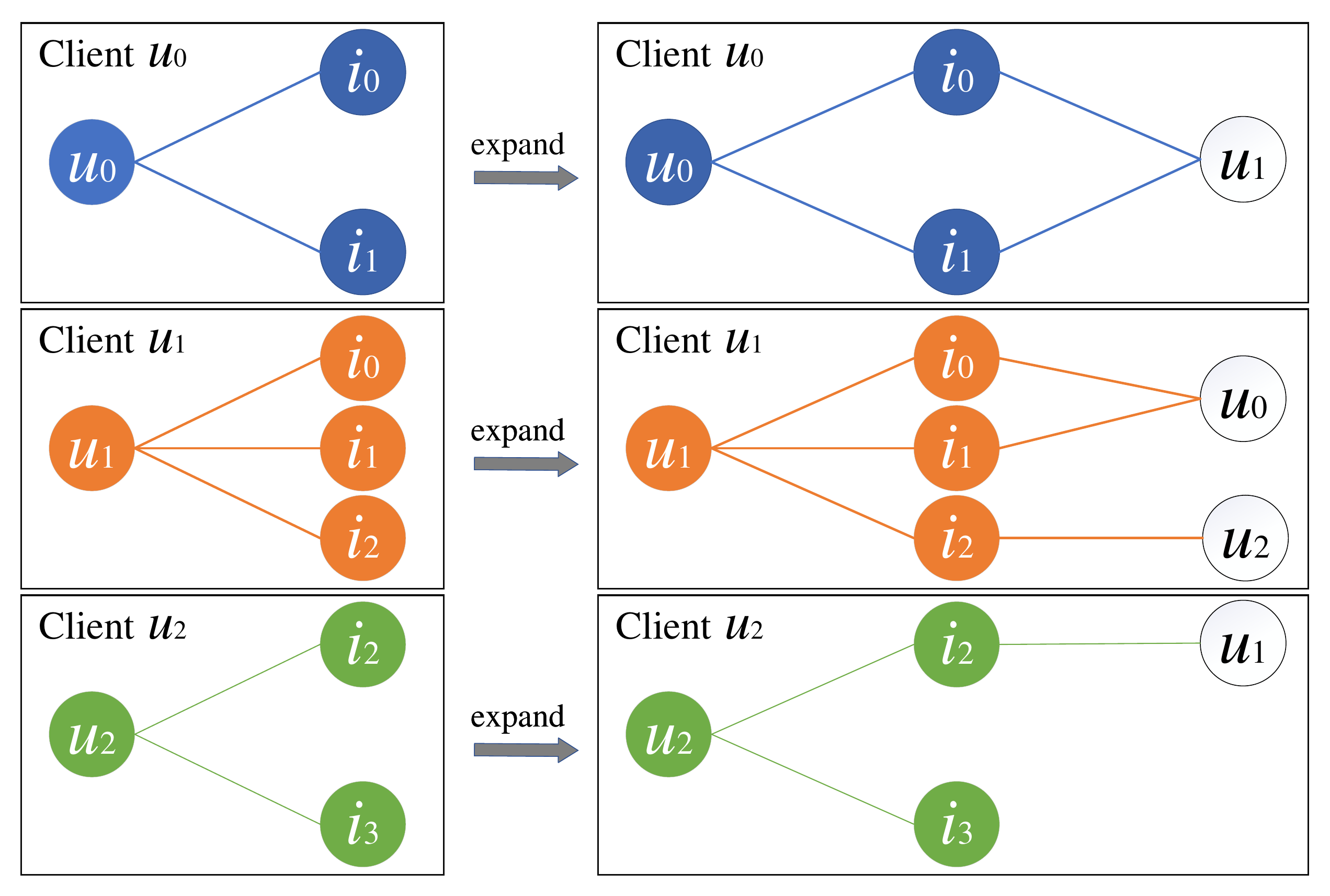}
\caption{Illustration of the process that each client expands the local subgraph in the example.}
\label{fig_expand_subgraph}
\end{minipage}%
}%
\subfigure{
\begin{minipage}[t]{0.49\linewidth}
\centering
\includegraphics[width=0.69\linewidth]{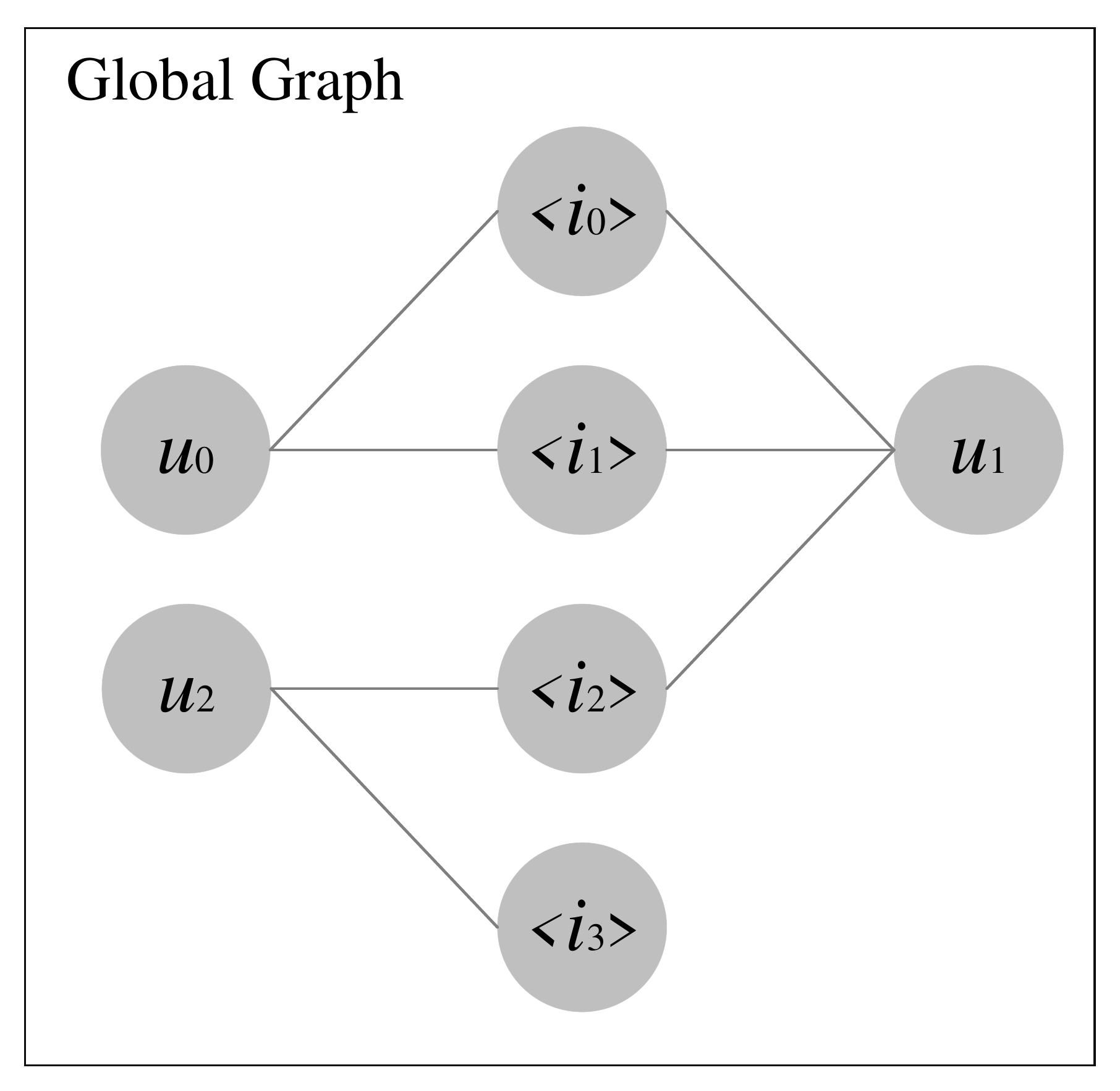}
\caption{Illustration of the process that the global graph constructed by the server. Notice that <$i_1$> means the ciphertext of $i_1$.}
\label{fig_global_graph}
\end{minipage}%
}%

\centering
%\caption{ pics}
\end{figure}

\begin{algorithm}[ht]
	\caption{ExpandLocalSubgraph}
        \label{Expand the local subgraph}
	\begin{algorithmic}[1]
		\FOR{each client $u$ in parallel}
		\STATE encrypts $\mathcal{I}_u$ (i.e., the set of item-IDs which are interacted by user $u$) with $S$
		\STATE sends <$\mathcal{I}_u$> (i.e., the cyphertext of $\mathcal{I}_u$) to the server
		\ENDFOR
		\\//Server
		\STATE receives the ciphertext from all clients	
		\STATE constructs the global graph by comparing the ciphertext
		\FOR{$u \in \mathcal{U}$}
		\STATE sends the neighboring users $\mathcal{N}_u$ and $\mathcal{I}^{e}_u$ (i.e., the set of items which are only interacted by user $u$) to the client $u$
		\STATE informs the client $u$ of the connectivity between neighboring users and common items
		\ENDFOR
		\FOR{each client $u$ in parallel}
		\STATE receives the information from the server
		\STATE expands the local subgraph
		\ENDFOR		
	\end{algorithmic}
\end{algorithm}
For expanding the local subgraph, we show the process in Algorithm~\ref{Expand the local subgraph}. Firstly, each client encrypts the ID of interacted items with the shared key $S$ and sends them to the server. Secondly, the server receives the ciphertext from all clients, then constructs the global graph by comparing them, which is shown in Figure~\ref{fig_global_graph}. For example, suppose there is a common item $i_0$ which is interacted with by two clients $u_0$ and $u_1$. Encrypting the same content with the same key will result in identical output. Therefore, the server knows that the clients $u_0$ and $u_1$ have a common item $i_0$, but could not know what item $i_0$ is due to just knowing the ciphertext of the item ID of $i_0$, which protects the privacy. Thirdly, for each client, the server sends the neighboring users-IDs and exclusive items (i.e., items that are only interacted with by the client, for instance, item $i_3$ is the exclusive item of client $u_2$) to them and informs the connectivity between neighboring users and common items. Finally, each client expands the local subgraph, as shown in Figure~\ref{fig_expand_subgraph}.\par 

\begin{algorithm}[ht]
	\caption{ForwardPropagation}
        \label{Train. forward propagation}
	\begin{algorithmic}[1]
		\FOR{ $ l=0, 1, \ldots, L-1$ }
		\FOR{each client $u$ in parallel}
		\STATE encrypts the user embedding of user $u$ in $l$-th layer $U^{[l]}_u$ with $S$
		\STATE sends them to the server
		\ENDFOR		
		\STATE the server receives the ciphertext from all clients and forwards them to the corresponding clients
		\FOR{each client $u$ in parallel}
		\STATE receives the ciphertext
		\STATE decrypts $U^{[l]}_u$ with $S$
		\STATE updates the $l$-th layer neighboring user embeddings in the local subgraph
		\STATE convolves the local subgraph at $l$-th layer to get the $(l+1)$-th layer user embedding and item embeddings(except for the neighboring users' embeddings)
		\ENDFOR	
		\ENDFOR		
	\end{algorithmic}
\end{algorithm}

\begin{figure*}
\centering
\subfigure{
\begin{minipage}[t]{0.49\linewidth}
\centering
\includegraphics[width=1.0\linewidth]{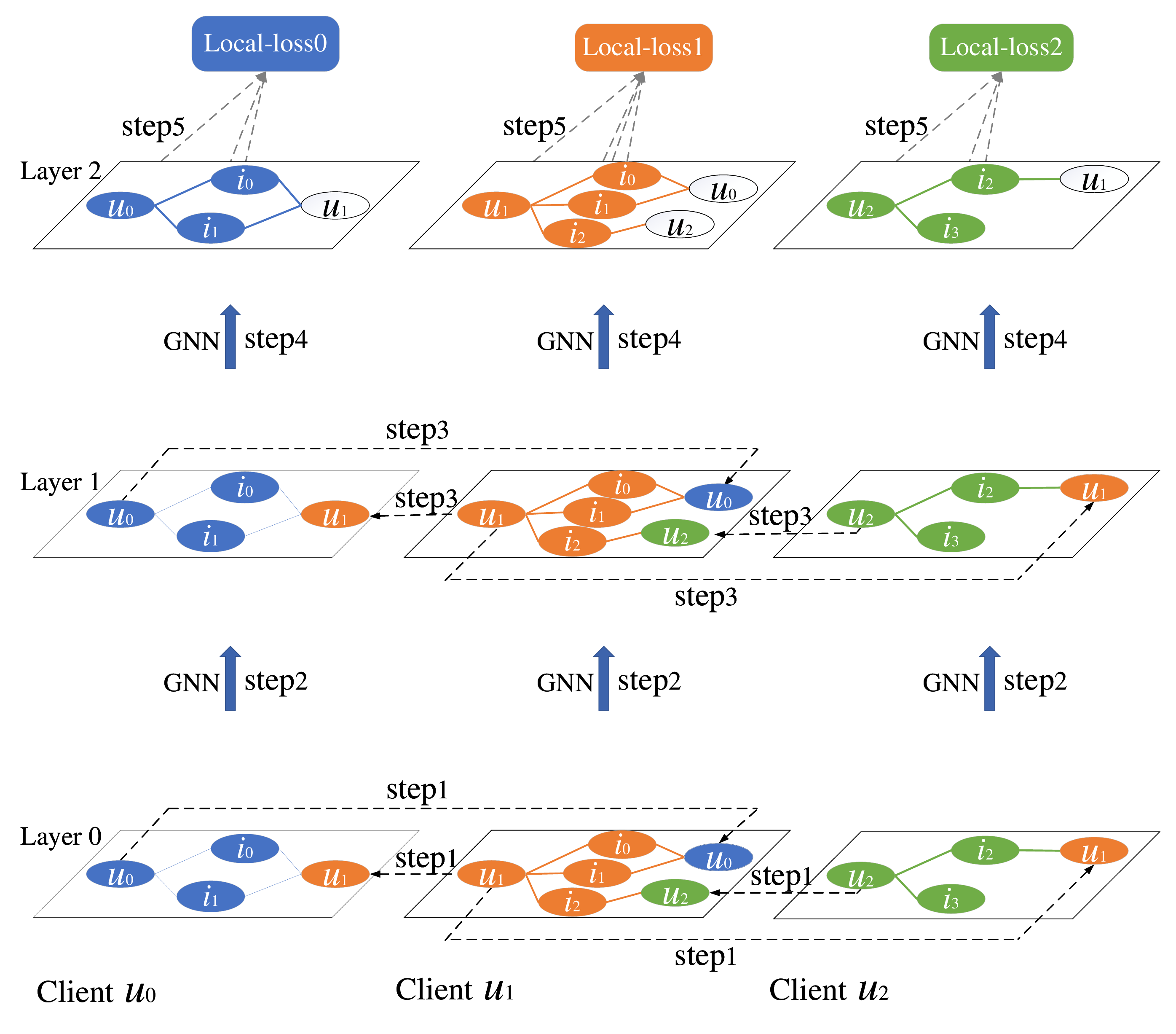}
\caption{Illustration of forward propagation.}
\label{fig_forward_propagaiton_new}
\end{minipage}%
}%
\subfigure{
\begin{minipage}[t]{0.49\linewidth}
\centering
\includegraphics[width=1.0\linewidth]{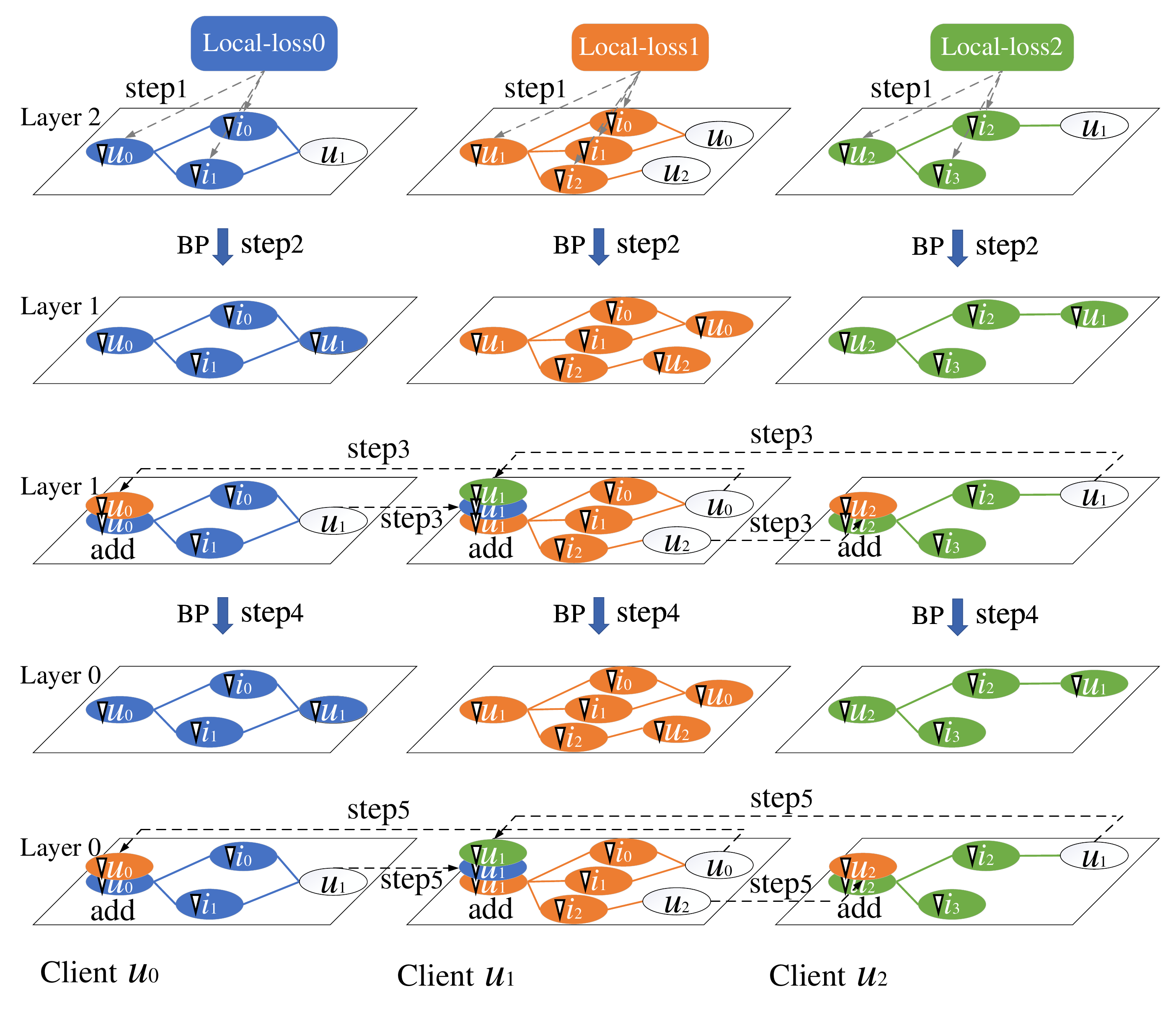}
\caption{Illustration of backward propagation.}
\label{fig_backward_propagation_new}
\end{minipage}%
}%

\centering
%\caption{ pics}
\end{figure*}

For forward propagation, we show the process in Algorithm~\ref{Train. forward propagation}. Firstly, synchronize users' embeddings. As shown in step 1 in Figure~\ref{fig_forward_propagaiton_new}, each client sends the user embedding to the neighboring users through the server. The transmission policy is that the sender encrypts and the receiver decrypts with the same shared key $S$. Since the server can receive the ciphertext in the process, but does not have the key to decrypt it, the privacy is protected. Secondly, each client convolves the local subgraph to get the user embedding and item embeddings of layer 1. Notice that we do not get the neighboring users’ embeddings by convolution, but receive them from the corresponding clients. i.e., step 3 in Figure~\ref{fig_forward_propagaiton_new}. Thirdly, each client convolves the local subgraph to get the user embedding and item embeddings of layer 2, i.e., the last layer.

\begin{algorithm}[ht]
	\caption{BackwardPropagation}
        \label{Train. backward propagation}
	\begin{algorithmic}[1]
		\FOR{ $ l=L,L-1, \ldots, 1$ }
		\FOR{each client $u$ in parallel}
		\STATE computes $\nabla^{[l-1]}_u$ with $\nabla^{[l]}_u$ (i.e., the gradient of $U^{[l]}_u$)
		\STATE encrypts $\nabla^{[l-1]}_{\mathcal{N}_u}$ (i.e., the gradient of $U^{[l]}_{\mathcal{N}_u}$) with $S$
		\STATE sends them to the server
		\ENDFOR		
		\STATE the server receives the ciphertext from all clients and forwards them to the corresponding clients
		\FOR{each client $u$ in parallel}
		\STATE receives the ciphertext
		\STATE decrypts them with $S$
		\STATE adds them to $\nabla^{[l-1]}_u$ (i.e., the gradient of all nodes w.r.t. user $u$ in the $l$-th layer)
		\ENDFOR	
		\ENDFOR		
	\end{algorithmic}
\end{algorithm}

For constructing the local loss and backward propagation, we show the process in Algorithm~\ref{Train. backward propagation}. Firstly, each client constructs a local loss with user embedding and item embeddings, which does not require the embedding of the neighboring users. Secondly, backward propagation. As shown in step 1 in Figure~\ref{fig_backward_propagation_new}, each client backpropagates to get the user embedding and item embedding gradients. Then backpropagate again to get the gradient of all nodes at layer 1. The gradient of the neighboring users would be sent back to the corresponding client, i.e., step 3 in Figure~\ref{fig_backward_propagation_new}. For example, the gradient of user embedding in client $u_1$ consists of three parts, one from client $u_0$, one from client $u_2$, and the final one from itself. Similarly, continue to backpropagate and finally get the embedding gradient of the user at layer 0 and the embedding gradient of the items.\par 

\begin{algorithm}[ht]
	\caption{AggregateGradients}
        \label{Train. aggregate the gradients}
	\begin{algorithmic}[1]
		\FOR{each client $u$ in parallel}
		\STATE split the gradient of item $i$ in $0$-th layer $\nabla^{[0]}_i$ $(i \in \{\mathcal{I}\backslash \mathcal{I}^e_u\})$ into two parts randomly
		\STATE chooses one of them to encrypt with $S$
		\STATE sends them to the server
		\ENDFOR		
		\STATE the server receives the ciphertext from all clients and forwards them to the corresponding clients
		\FOR{each client $u$ in parallel}
		\STATE receives the ciphertext of the gradients of items
		\STATE decrypts them with $S$
		\STATE adds them to the corresponding node of items
		\STATE uploads $\nabla^{[0]}_i$ $(i \in \{\mathcal{I}\backslash \mathcal{I}^e_u\})$ to the server
		\ENDFOR	
		\STATE the server receives the gradients of items from all clients, aggregates them and distributes them to the corresponding clients
	\end{algorithmic}
\end{algorithm}
\begin{figure}
\centering
\subfigure{
\begin{minipage}[t]{0.49\linewidth}
\centering
\includegraphics[width=1.0\linewidth]{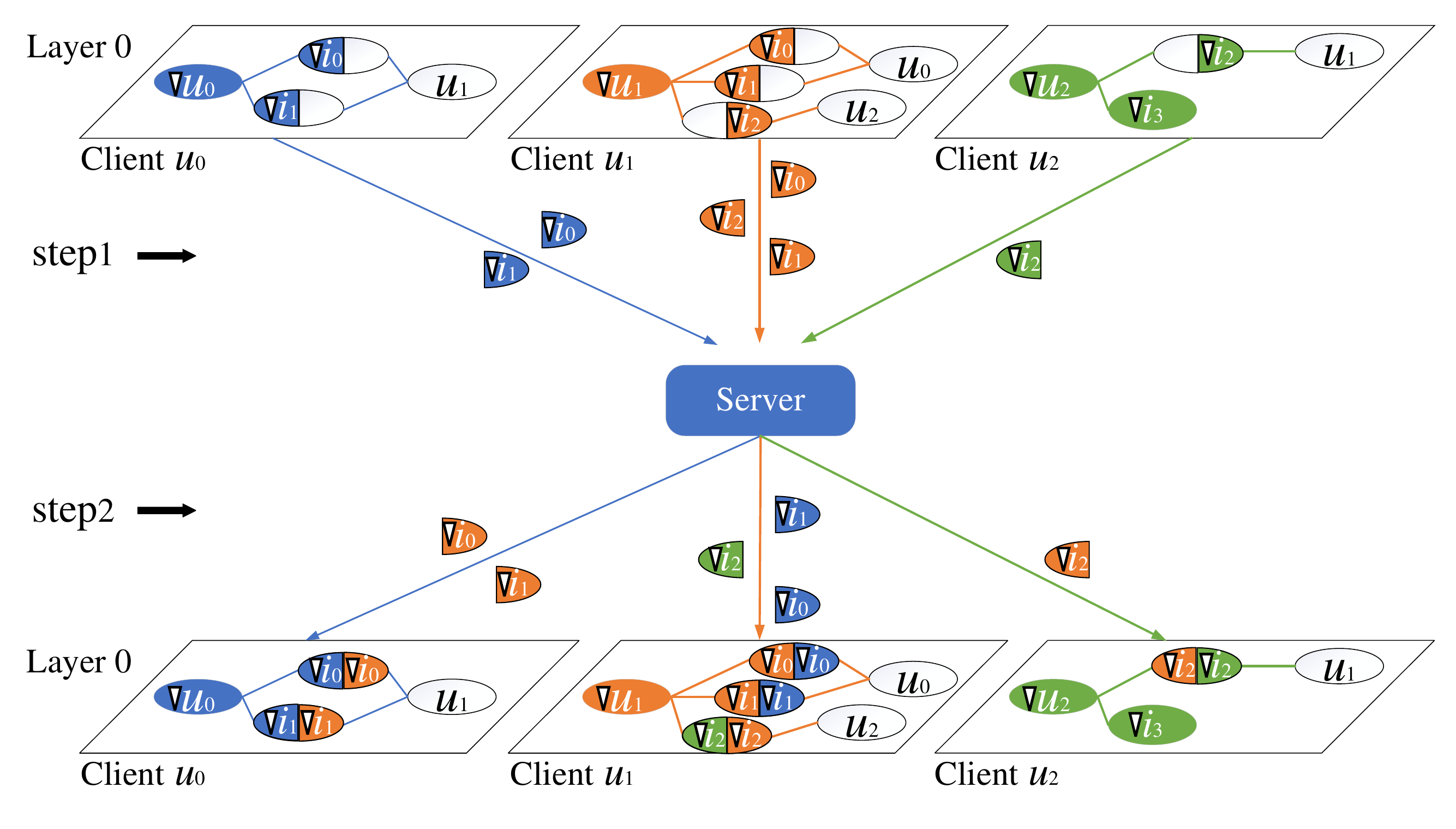}
\caption{Illustration of sending one part of gradients.}
\label{fig_aggregate_gradients_step1}
\end{minipage}%
}%
\subfigure{
\begin{minipage}[t]{0.49\linewidth}
\centering
\includegraphics[width=1.0\linewidth]{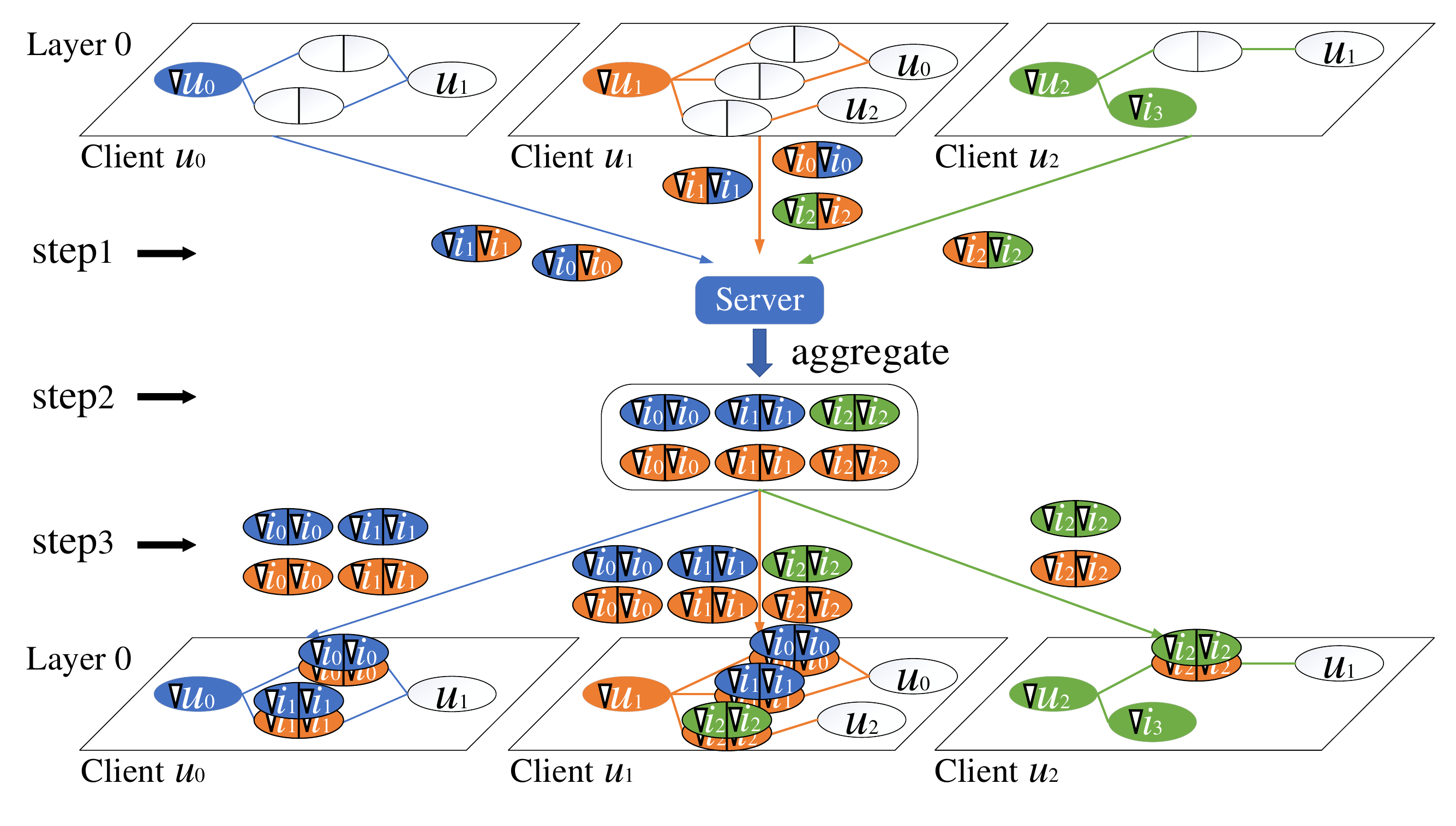}
\caption{Illustration of aggregating gradients.}
\label{fig_aggregate_gradients_step2}
\end{minipage}%
}%

\centering
%\caption{ pics}
\end{figure}

As for how to aggregate the gradients, we show the process in Algorithm~\ref{Train. aggregate the gradients}. Notice that only the gradients of the item embeddings (except the exclusive items and user embedding of each client) need to be aggregated. Here we use the secret sharing technology~\cite{shamir1979share} to protect privacy. Firstly, each client splits the aggregated gradients into two parts randomly, chooses one to encrypt with the shared key S, and sends them to the one of neighboring clients by the server, as shown in Figure ~\ref{fig_aggregate_gradients_step1}. Secondly, each client decrypts the ciphertext and adds them to the corresponding nodes. Finally, as shown in Figure~\ref{fig_aggregate_gradients_step2}, each client sends the gradients to the server for aggregation.\par 

\begin{algorithm}[ht]
	\caption{Predict}
        \label{Predict}
	\begin{algorithmic}[1]
		\FOR{each client $u$ in $\mathcal{U}^{,}_I$ (i.e., the set of $u^,_i$, $i \in \mathcal{I}$) ($u^{,}_i$ means the user randomly chosen by the server in the set of users who interact with item $i$)}
		\STATE encrypts $V_u$ (i.e., the set of item embeddings that need to be uploaded by the client $u$) with $S$
		\STATE sends them to the server
		\ENDFOR		
		\STATE the server collects the ciphertext of all item embeddings and then sends the negative item embeddings (i.e., the items which are not interacted with by the user) to all clients
		\FOR{each client $u$ in parallel}
		\STATE receives the ciphertext of negative item embeddings
		\STATE decrypts them with the shared key $S$
		\STATE calculates the predicted scores with user embedding and item embeddings
		\ENDFOR	
	\end{algorithmic}
\end{algorithm}

We now introduce how to make a prediction (i.e., Algorithm~\ref{Predict}). Firstly, for each item, the server randomly selects one user from those who have interacted with it to send its embedding encrypted with the shared key S. Secondly, the server collects the ciphertext of all item embeddings. Then for each client, the server sends the embeddings of the negative items, which refer to the items that the client has not previously interacted with. Finally, each client calculates the predicted scores with user embedding and item embeddings.
\section{EXPERIMENTAL SETTING}
\subsection{Dataset and Evaluation Metrics}
We use the files u1.base and u1.test of MovieLens 100K as our training data and test data, respectively. MovieLens 100K contains 943 users and 1682 items. The u1.base file contains 80000 rating records, and its density is 5.04\%. The u1.test file contains 20000 rating records. We follow the common practice and keep the (user, item) pairs with ratings 4 or 5 in u1.base and u1.test as preferred (user, item) pairs, and remove all other records. \par
We use two commonly used evaluation metrics, i.e., Precision@N and Recall@N, where N is the number of recommended items. 

% The source code can be found at~\href{https://anonymous.4open.science/r/GNN4FR-code-ED93}{https://anonymous.4open.science/r/GNN4FR-code-ED93}
\subsection{Results}
We use LightGCN to instantiate an example of our framework and report the results in Table~\ref{experiment_result}. The training process and test results are equivalent to the corresponding un-federated counterpart. Notice that we do not include more datasets and baselines because the purpose is to show the equivalence between the proposed framework and the un-federated counterpart, which is different from that of empirical studies in most works.

\begin{table}[hbp]
 \centering
 \caption{Experimental results of the proposed federated method GNN4FR and its un-federated version LightGCN.}
 \label{experiment_result}
 \begin{tabular}{ccc}
  \toprule
  Model & Precision@5 & Recall@5  \\
  \midrule
  LightGCN  & 0.3816 & 0.1257 \\
  GNN4FR     & 0.3816 & 0.1257 \\

  \bottomrule
 \end{tabular}
\end{table}

\section{Conclusions and Future Work}
In this paper, we propose the first lossless GNN-based federated recommendation framework named GNN4FR, which uses an “expanding local subgraph + synchronizing user embedding” mechanism to achieve full-graph training, enabling the training process to be equivalent to the corresponding un-federated approach. In addition, we leverage secret sharing to protect privacy while aggregating the gradients. Finally, we use LightGCN to instantiate an example of our framework and show its feasibility and equivalence. For future work, we aim to develop diverse models using GNN4FR that can accommodate specific algorithms while achieving a good balance between accuracy and efficiency.

\bibliographystyle{ACM-Reference-Format}
\bibliography{sample-authordraft}

\end{document}